\documentclass[10pt]{article}
\usepackage[a4paper]{geometry}

\usepackage{graphicx} 
\usepackage{hyperref}
\usepackage{booktabs}
\usepackage{placeins}
\usepackage{algorithm}
\usepackage{algpseudocode}
\usepackage{amsmath,amssymb, amsthm}
\usepackage{amsfonts}
\usepackage{cleveref}
\usepackage{xcolor}
\usepackage[labelfont=bf]{caption}
\usepackage{subcaption}
\usepackage{parskip}
\usepackage{minibox}
\usepackage{authblk}

 \newtheorem{theorem}{Theorem}
  \newtheorem{lemma}[theorem]{Lemma}
 \newtheorem{property}[theorem]{Property}
 \newtheorem{fact}[theorem]{Fact}
  \newtheorem{corollary}[theorem]{Corollary}
  \newtheorem{conjecture}[theorem]{Conjecture}

\newcommand{\R}{\mathbb{R}}

\newcommand{\eps}{\varepsilon}
\newcommand{\coreset}{\Omega}
\newcommand{\cost}{\textsc{cost}}
\newcommand{\opt}{\textsc{opt}}
\newcommand{\E}{\mathbb{E}}
\newcommand{\coresetSize}{n_c}

\DeclareMathOperator*{\dist}{dist}
\DeclareMathOperator*{\polylog}{polylog}
\DeclareMathOperator*{\poly}{poly}

\newcommand{\lpar}{\left(}
\newcommand{\rpar}{\right)}
\newcommand{\lbra}{\left\{}
\newcommand{\rbra}{\right\}}

\newcommand{\calA}{\mathcal{A}}

\newcommand{\calE}{\mathcal{E}}

\usepackage[colorinlistoftodos,prependcaption,textsize=tiny]{todonotes}

\usepackage{mdframed}
\usepackage{wrapfig}
\newcommand{\erclogowrapped}[1]{%
\setlength\intextsep{0pt}%
\begin{wrapfigure}[3]{r}{#1*\real{1.1}}%
\includegraphics[width=#1]{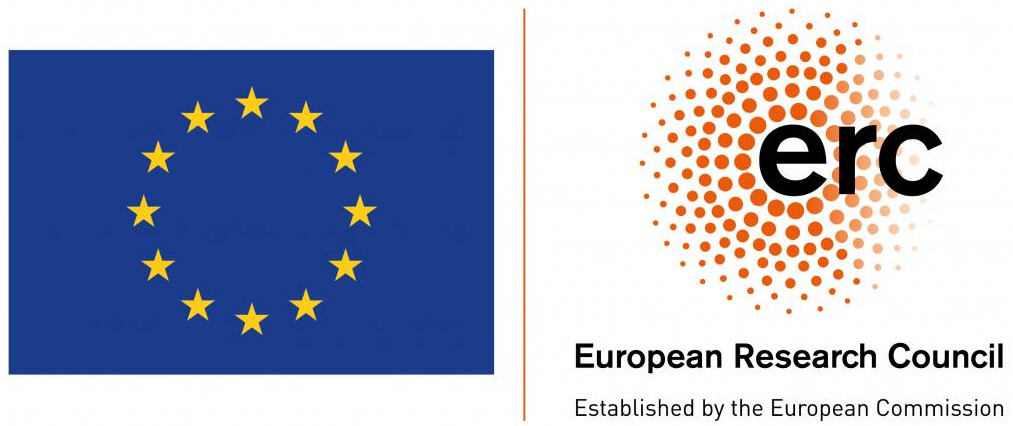}%
\end{wrapfigure}%
}

\title{Fully Dynamic $k$-Means Coreset in Near-Optimal Update Time}

\date{}

\author[1]{Max Dupré la Tour}
\author[2]{Monika Henzinger}
\author[3]{David Saulpic}

\affil[1]{McGill University, Montreal, Canada}
\affil[2]{Institute for Science and Technology Austria (ISTA), Klosterneuburg, Austria}
\affil[3]{Université Paris Cité \& CNRS, IRIF, Paris, France}

\begin{document}
\maketitle

\begin{abstract}
    We study in this paper the problem of maintaining a solution to $k$-median and $k$-means clustering in a fully dynamic setting. 
    To do so, we present an algorithm to efficiently maintain a coreset, a compressed version of the dataset, that allows easy computation of a clustering solution at query time. Our coreset algorithm has near-optimal update time of $\tilde O(k)$ in general metric spaces, which reduces to $\tilde O(d)$ in the Euclidean space $\R^d$. The query time is $O(k^2)$ in general metrics, and $O(kd)$ in $\R^d$. 
    
    To maintain a constant-factor approximation for $k$-median and $k$-means clustering in Euclidean space, this directly leads to an algorithm update time $\tilde O(d)$, and query time $\tilde O(kd + k^2)$. To maintain a $O(\polylog k)$-approximation, the query time is reduced to $\tilde O(kd)$.
\end{abstract}

\section{Introduction}
As a staple of data analysis, the problem of clustering a dataset has been widely investigated, both from a theory and practice perspective. In this paper, we focus on the fully dynamic setting, where the input changes through insertions and deletions of data items: as an illustration, we think of a data-miner clustering their dataset, analyzing it, and suitably modifying it -- for instance, to clean the dataset by removing outliers, or by adding new observations. In the big-data setting, updates need to be very efficient: instead of recomputing a solution from scratch, the goal of fully dynamic clustering algorithms is to efficiently cluster the data after a user updates.
This problem has received a lot of attention under various models of clustering: recent works  tackled the $k$-center problem~\cite{ChanGS18, BateniEFHJMW23}, effectively closing the problem from the theory side; the facility location problem~\cite{Cohen-AddadHPSS19}; and the more intricate $k$-median and $k$-means problems~\cite{Cohen-AddadHPSS19, HenzingerK20, sayan23}. The latter is probably the most useful for practice as $k$-means is one of the most common unsupervised learning techniques, and this paper presents new improvements for both $k$-median and $k$-means problems. 
The input for the $k$-median problem is a set of weighted points in some metric space, and the goal is to output a set of $k$ centers that minimizes the sum, for each input point, of its weight times its distance to the closest center. For $k$-means, it is the sum of squared distances. 
Both problems are NP-hard to approximate arbitrarily close to one~\cite{GuK99}: therefore, we focus on finding a set of centers with cost (i.e., sum of distances) within a constant factor of the optimal solution.
 
A standard  approach to solve $k$-median and $k$-means problems is the use of \emph{coreset}. An $\eps$-coreset for input $P$ is a compressed version of the input, $\coreset$, such that for \emph{any} set of $k$ centers $S$ the cost is almost the same when it is evaluated on the full input or on the coreset, i.e.: 
$$\cost(\coreset, S) = (1\pm \eps)\cost(P, S).$$
Perhaps surprisingly, coresets can have size \emph{almost independent of $n$}: the smallest $\eps$-coresets in general metric spaces have size $O(k\log(n) \eps^{-2})$, which can be replaced with $O(k \eps^{-4})$ in Euclidean spaces~\cite{stoc21}.
As the cost of any candidate solution of $k$ centers is preserved, an algorithm that dynamically maintains a coreset can therefore be directly used to maintain dynamically a solution to $k$-means: at query time, simply use a \textit{static} approximation algorithm on the coreset. 
Therefore, using a $k$-means algorithm with running time $T(n,k)$ on a dataset of size $n$, this translates into a fully dynamic algorithm with query time $T(O(k\log(n)), k)$ (and an update time that depends on the coreset algorithm itself).

To maintain a coreset in a dynamic setting, a \emph{merge-and-reduce} procedure was introduced by Har-Peled and Mazumdar \cite{Har-PeledM04} 
(based on Bentley and Saxe \cite{BentleyS80}
and Agarwal, Har-Peled, and Varadarajan~\cite{agarwal2004approximating}).
This transforms any static coreset construction into a fully-dynamic one, reducing the dependency in $n$ to a dependency in $O(k\log(n))$ -- the same speed-up as when applying an approximation algorithm on a coreset. With current coreset constructions, this translates into an algorithm with worst-case  query and {\em amortized}  update time $\tilde O(k^2)$.\footnote{We use the notation $\tilde O(T)$ as a shorthand to $O(T \polylog(n))$.} Henzinger and Kale~\cite{HenzingerK20} showed how to solve the problem in $\tilde O(k^2)$ {\em worst-case} query and update time.
The interest of this method is two-fold: first, it uses a static coreset algorithm in a  black-box manner. Therefore, in a setting where faster coreset algorithms exist (such as Euclidean space), the algorithm  directly improves. Second, at query time, the algorithm uses any static $k$-means algorithm: here as well, a different algorithm may be used to achieve a different time-to-accuracy ratio. 

To improve the update time, Bhattacharya, Costa, Lattanzi and Parotsidis~\cite{sayan23} recently adopted a completely different strategy: they adapted an old and fast algorithm from Mettu and Plaxton \cite{MettuP04}, and showed how to implement it such that only a few changes are required between each update. That way, they manage to improve the amortized update time to $\tilde O(k)$ with $\tilde O(k^2)$ query time (which is near-optimal for any $O(1)$-approximation algorithm in general metric space according to a lower bound in~\cite{MettuP04,sayan23}), but with an (unspecified) constant approximation ratio. 
However, their result is much more restrictive than the coreset framework: (a) it only works in general metric spaces and cannot be fine-tuned to specific cases, and (b) it is based on an approximation algorithm with an unspecified constant approximation factor.
Furthermore, they state in their introduction that it is ``not at all clear how to use [the coreset] algorithm to reduce the update time to $\tilde O(k)$". This appears true when  a static coreset algorithm is used as a black-box in the framework of \cite{Har-PeledM04,HenzingerK20}, but possibly not in general: in this work, we therefore ask
\begin{center}
        \minibox[c, frame]{ Is it possible to design a  coreset construction suited specifically to dynamic clustering algorithms?}
\end{center}

We answer  this question positively, showing how to use the framework of \cite{Har-PeledM04} with a particular coreset construction -- answering the (implicit) question from \cite{sayan23}. 
One key advantage of our technique is that, if the metric space of study admits some faster $k$-means algorithm, then our algorithm is faster as well. In particular, applied to Euclidean space, we get \emph{almost constant} update-time, and \emph{almost linear} query time.

\subsection{Our result and techniques}

More precisely, we show the following theorem.
\begin{theorem}\label{thm:main}
    There exists an algorithm for fully dynamic $k$-median (resp. $k$-means), that maintains an $\eps$-coreset of size $\tilde O\lpar k \eps^{-2}\rpar$ with amortized update and query time $O\lpar \frac{T(k \polylog(n), k)}{k}\rpar$, where $T(k \polylog(n), k)$ is the time to compute a $\polylog(n)$-approximation to $k$-median (resp. $k$-means) on a dataset of size $n$.
\end{theorem}

As mentioned above, we can combine this algorithm with \emph{any} static $k$-median (resp. $k$-means) algorithm: if the static algorithm runs in time $T'(n, k)$ on a dataset of size $n$ and computes a $c$-approximation, then there is a fully-dynamic algorithm to solve $k$-median (resp. $k$-means) with amortized update time $O\lpar \frac{T(k \polylog(n), k)}{k}\rpar$, query time $T'(k \eps^{-2} \polylog(n), k)$ and approximation guarantee $2(1+\eps)c$ (resp. $4(1+\eps)c$).\footnote{The factor $2$ and $4$ come from the fact that, to achieve a fast query time, we restrict the set of potential centers to the coreset points. See \cite{HenzingerK20} for precise computation.}

\begin{corollary}
In general metric spaces, there is a coreset-based fully-dynamic algorithm to compute an $O(1)$-approximation to $k$-median (resp. $k$-means) with amortized update time $\tilde O(k)$ and worst-case query time $\tilde O(k^2)$.
In $d$-dimensional Euclidean space, the amortized update time is $\tilde O(d)$, and the worst-case query time is $\tilde O(k^2 + kd)$.
\end{corollary}
Note that the running-time for general metric spaces matches the results of \cite{sayan23}, but we improve the constant in the approximation ratio\footnote{They adapt a static $O(1)$ algorithm, and making it dynamic blows up the approximation ratio by an unspecified $O(1)$ factor. On the other hand, our algorithm preserves the approximation ratio of any static algorithm, up to a multiplicative factor $4(1+\epsilon)$.}: our algorithm matches the approximation ratio in~\cite{Har-PeledM04,HenzingerK20}, while improving their update time.
In Euclidean space, the query-time of our corollary matches the one from \cite{HenzingerK20, sayan23}, but we reduce the update-time from $\tilde O(k^2d)$ to $\tilde O(kd)$. Furthermore, since it is possible to compute an $O(\log k)$-approximation to the static problems in time $O(nd)$ (see \cref{sec:euclideanSpace}), our algorithm can achieve this approximation factor with query time $\tilde O(kd)$, which equals the time to output the solution, namely a list of $k$ centers in $\R^d$. 
We conjecture that an $O(1)$-approximation can be computed with the same running time, which would provide a likely near-optimal algorithm for both update and queries; we discuss this more in \Cref{sec:euclideanSpace}.

\textbf{Our techniques.}
It is elementary to show our theorem in insertion-only streams. Indeed, a key property of coreset ensures that if $\coreset$ is a coreset for $P$, $\coreset \cup \{x\}$ is a coreset for $P \cup \{x\}$.  Therefore, if the goal is to maintain a coreset of size $\tilde O(k)$ under insertions, one can merely compute a coreset every $k$ steps, in time $T(n,k)$ (the same time as computing an $\tilde O(1)$-approximation), and in-between simply add new points to the coreset in time $O(1)$. 
Using this idea in the merge-and-reduce algorithm of \cite{Har-PeledM04} directly yields an amortized runtime of essentially $T(k, k)/k$ -- which is $\tilde O(k)$ in general metric spaces, $\tilde O(d)$ in $\R^d$.

Dealing with deletions is more intricate as it is \textit{not} the case that $\coreset \setminus \{x\}$ is a coreset for $P \setminus \{x\}$\footnote{Consider the case where $x$ concentrates most of the cost of a solution: because of the multiplicative approximation guarantee, $\coreset$ only needs to capture the cost of $x$, and have a poor approximation elsewhere. Removing $x$ requires the coreset to be precise everywhere else.}.
Standard coreset constructions are randomized, and the coreset is built by sampling according to an intricate sampling distribution, typically \emph{sensitivity sampling}. This sampling method works as follows: start by computing a good approximation to $k$-means (with possibly $O(k)$ centers), and then sample points proportionally to their cost in the solution, and weight each point inversely to their sampling probability.
There are two difficulties in adapting such a sampling algorithm. (1) The initial good approximation to $k$-means does not necessarily remain good after a deletion -- a single deletion can make the approximation ratio go from 1 to infinity\footnote{In a set of $k+1$ points at distance $1$ from each other any set of $k$ distinct centers is an optimal solution with cost $1$. However after removing a single point (which is a center) the cost of the optimal solution is $0$ while the solution still has cost $1$.}
(2) Even without changing the solution, maintaining efficiently a sample proportionally to its contribution to the solution is not obvious at all.

We address those two questions independently. (1) We show that a good solution for $2k$-means remains a good solution for $k$-means, even after $k$ deletions. 
This allows to compute a solution only every $k$ steps -- just as in the insertion only case. This result is novel in the world of dynamic clustering algorithms: it is the first result that allows to build some approximation and let it deteriorate over time, while still being a ``good enough'' solution.

(2) We show how to build a coreset from \textit{uniform sampling} instead of sampling from a more complicated  distribution. 
In this way it is much easier to maintain a uniform sample under insertions and deletions: we build on the coreset construction of \cite{stoc21}, to show that, if the input satisfies the following property, called \Cref{prop:groups}, any  small uniform sample is a coreset.
\begin{property}\label{prop:groups}
    Let $G$ be a set of points with weights in $[1, 2]$, and  let $A$ be a set of centers.  We say $G$ and $A$ satisfy \Cref{prop:groups} when:
    \begin{itemize}
\item Each point has the same cost up to a factor $2$, namely, $\forall p, q \in G, \cost(p, A) \leq 2 \cost(q, A)$.
\item All non-empty clusters in solution $A$ have the same size, up to a factor $8$: there is a scalar $c_A$ such that, for any cluster $C$ in solution $A$, either $C \cap G = \emptyset$ or $c_A \leq \cost(C \cap G, A) \leq 8 c_A$.
\end{itemize}
\end{property} 
In the above, $A$ will be a set of $O(k)$ centers; and for each center $a \in A$, the cluster of $a$ is the set of points closer to $a$ than any other center of $A$.
Therefore, we need to (1) maintain a small number of groups of data points with the property above, and (2) maintain a uniform sample in each group. We show that, for $k$ consecutive steps, our algorithm does so very efficiently.

Combining those results allows us to use the same strategy for deletions as for insertions: compute a fresh solution every $k$ steps, and update it very efficiently between two recomputations. Therefore, the amortized running time of the algorithm is $T(n, k)/k$, where $T(n, k)$ is the running time to compute a coreset for a dataset of size $n$ (or, equivalently, an $\tilde O(1)$-approximate solution). 
Finally, we incorporate this into the merge-and-reduce framework of \cite{Har-PeledM04}, to reduce further the running time to $T(k, k)/k$. 
This is the most intricate part of our analysis: for a reason that will later become obvious, we need to ensure that the coreset algorithm produces a coreset that evolves \emph{not more} than the input: when a single point is deleted, a single point should be deleted from the coreset. This is formalized in the following theorem, which is our main technical contribution:
\begin{lemma}\label{lem:singleCoreset}
    For $0 < \eps \leq 1/3$, there exists an algorithm that maintains an $\eps$-coreset under insertion and deletions, with the following guarantee. Starting with a dataset $P_0 \subset \R^d$ of size $n$, the running time for initializing the data structure is $\tilde O(T(n, 2k))$, where $T(n, k)$ is the running time to compute an $\tilde O(1)$-approximation for $k$-means on a dataset of size $n$. 
    Then, after $n_i$ insertions and $n_d$ deletions with $n_i+n_d \leq k$, it holds that: (1) the total running time for updates is at most $\tilde O(n_i+n_d)$, (2) the total number of points inserted into the coreset is at most $n_i+n_d$, and (3) the total number of points deleted from the coreset is at most $n_d$.
\end{lemma}

Those ideas allow us to maintain efficiently a precise coreset. In order to answer a $k$-means query, we simply use a static algorithm on the coreset.

\subsection{Further related work}
We already covered the closest related work on dynamic $k$-median and $k$-means clustering. 
In the particular case of $k$-median in Euclidean spaces, where points are restricted to the grid $\{0, 1, ..., \Delta\}^d$, \cite{BravermanFLSY17} achieve update time $O(d \log^2 \Delta)$. This, however, only works for $k$-median, because of the use of a quadtree embedding. 
For $k$-means, the preprint \cite{hu2018nearly} claims to show how to maintain an $\eps$-coreset of size $\tilde O(k\eps^{-2} d^4 \log^2 \Delta)$, using at most $k \poly(d/\eps \cdot \log \Delta)$ bits of memory. We believe that the (un-specified) running time is the same as ours; however, the coreset computed is correct at any step with probability $0.97$, while ours is $1-\delta$ for any $\delta > 0$. This means that their algorithm fails every $\approx$ 30 queries, and we do not see an immediate fix to this.\footnote{In particular, note that the problem of estimating whether a given set is an $\eps$-coreset is co-NP hard \cite{SchwiegelshohnS22}.}

To maintain coresets in general metric spaces, \cite{Har-PeledM04} introduced the merge-and-reduce tree with an amortized running-time analysis, and \cite{HenzingerK20} showed how to turn it into a worst-case guarantee.

The literature on coreset recently boomed (see e.g. \cite{FeldmanSS20, HuangV20, BravermanJKW21} and the references therein), with a series of work achieving optimal bound for $k$-means clustering of $\tilde O(k \eps^{-2} \log n)$ in general metric spaces \cite{stoc21} and $\tilde O(k\eps^{-4})$ or $\tilde O(k^{3/2} \eps^{-2})$ \cite{stoc21, Cohen-AddadLSSS22, huang23optimal}. 
Besides their use in the dynamic setting, coresets are key to a recent breakthrough in the streaming model \cite{Cohen-AddadWZ23}, which shows that a memory of only $dk \eps^{-4} \poly(\log \log n)$ is necessary.

The related $k$-center clustering is perhaps easier to handle: the reason is that a certificate that the cost is higher than a threshold only needs $k+1$ points (at distance more than  twice the threshold from each other). This is the basis of the works of \cite{ChanGS18, BateniEFHJMW23}. This can be more easily maintained than a $k$-median solution, for which no such certificate exists.

\subsection{Definitions and notations.}
The $(k,z)$-clustering problem is defined as follows. Given a set of points $P$ in a metric space $(X, \dist)$, with weights $w : P \rightarrow \R^+$, the goal is to find a set $S$ of $k$ points that minimizes the cost function $\cost_z(P, S) := \sum_{p \in P} w(p) \min_{s \in S} \dist(p, s)^z$. We call $S$ with $k$ points of $X$ a candidate solution. For $c\in \R$, a candidate solution $S$ is a $c$-approximation if $\cost_z(P, S) \leq c \min \cost_z(P, S')$, where the minimum is taken over all candidate solutions $S'$.

An $\eps$-coreset for $(k,z)$-clustering is a weighted set $\coreset$ such that, for any candidate solution $S$, $\cost_z(\coreset, S) = (1\pm \eps)\cost_z(P, S)$. 
In the following, we will use $\coresetSize$ to denote the size of the coreset, in particular the coreset constructed via \Cref{lem:coresetStatic}.

We use $T(n, k)$ to denote the running time of an algorithm computing an $\tilde O(1)$-approximation for $k$-means, on a dataset of size $n$. In Euclidean space $\R^d$, this is $\tilde O(nd)$. In general metric spaces, this is $O(nk)$.
We will sometimes abusively denote $T$ for $T(k \polylog(n), k)$. We use the notation $\tilde O(x)$ to denote $x \cdot \polylog(nx)$.

For simplicity of presentation, we will focus our presentation on $k$-means -- i.e., $(k,2)$-clustering -- and write $\cost$ for $\cost_2$. All the results can be directly translated to the more general problem, replacing $T(n,k)$ by the running time to compute an approximation to $(k,z)$-clustering.

\section{Preliminary results}

In this section, we provide some results that are crucial to our analysis, and can be used in a black-box manner. 
Our first lemma formalizes that, in a group of points that satisfy \Cref{prop:groups}, a uniform sample produces a coreset: 
\begin{lemma}
\label{lem:coresetStatic}
Let $G$ be a group of points with weights $w: G \rightarrow [1, 2]$ and let $A$ be a set of centers such that $G$ and $A$ satisfy \Cref{prop:groups}. 
Let $\coreset$ be a set of $\coresetSize = O\lpar k\log(n) \eps^{-2} \polylog(k/(\delta\eps))\rpar$ points sampled uniformly at random, where $p \in \coreset$ has weight $w(p) \cdot |G|/\coresetSize$. It holds with probability $1-1/\delta$ that  
\[\forall S, |\cost(\coreset, S) - \cost(G, S)| \leq \eps (\cost(G, A) + \cost(G, S)).\]
Furthermore, the total weight verifies: $\sum_{p \in G} w(p) = (1\pm \eps) \sum_{p \in \coreset} w(p) |G|/\coresetSize$.

In Euclidean space $\R^d$, it is enough to take $\coresetSize = O\lpar k \eps^{-4} \polylog(k/(\delta\eps))\rpar$.
\end{lemma}

This lemma is very related to Lemma 2 in \cite{stoc21}, which uses \emph{group sampling} to construct a coreset. 
This algorithm starts by computing a solution $\calA$, with set of clusters $A_1, ..., A_k$ and partitions the input into $\polylog(k, 1/\eps)$ structured groups. The groups have the following property:  each cluster that intersects with the group has roughly the same contribution to the cost of the group in the solution $\calA$, and each point within the same cluster has roughly the same distance to the center $\calA$. Then, the algorithm samples essentially $k \eps^{-4}$ points from each group, following the following distribution: in group $G_i$, point $p$ in some cluster $A_j$ is sampled with probability $w(p) \cdot \frac{\cost(A_j \cap G_i, \calA)}{|A_j \cap G_i| \cost(G_i, \calA)}$.

\Cref{prop:groups} is a strong requirement on $A$ designed such that the distribution of group sampling becomes essentially uniform: for a set of points satisfying \Cref{prop:groups}, the probability of sampling any two points (using the group sampling distribution) differs by a small constant factor. Therefore, it is not a surprise that the proof can be adapted to work with uniform sampling instead.
More precisely, group sampling will sample each point with probability within a constant factor of uniform. 
The analysis of group sampling is based on a concentration inequality on the sum of the random variables indicating whether each point is sampled or not. This concentration is based on bounding the first and second moments of those variables: since the group sampling distribution is close to the uniform sampling one, those moments are essentially the same and the proof goes through. We provide a thorough proof in \Cref{sec:uniformCoreset}. 
In the following, we will use $\coresetSize$ to denote the size of the coreset constructed via \Cref{lem:coresetStatic}.

As explained in the introduction, we crucially need to construct a bicriteria solution whose cost stays close to optimal over many steps. This can be achieved using the following result. 
\begin{lemma}\label{lem:bicriteria}
Let $A$ be a $c$-approximation to the $2k$-means problem on a weighted set $P$. Then, for any set $D \subset P, |D| \leq k$, it holds that $$\cost(P \setminus D, A) \leq c \cdot \opt_k(P \setminus D),$$ where $\opt_k(P \setminus D)$ is the optimal $k$-means solution on $P\setminus D$.
\end{lemma}
\begin{proof}
    We let $\opt_k(P \setminus D)$ be the optimal $k$-means solution on $P\setminus D$, and  $\opt_{2k}(P)$ be the optimal $2k$-means solution on $P$. 
    By definition of $A$, it holds that 
        $\cost(P \setminus D, A) \leq \cost(P, A) \leq c \cdot \cost(P, \opt_{2k}(P))$.
    Since $|\opt_k(P \setminus D) \cup D| \leq 2k$,
    $\opt_k(P \setminus D) \cup D$ is a $2k$-means solution for $P$ with cost at least as high as the cost of $\opt_{2k}(P)$.
    Thus, $A$ is a $2k$-means solution for $P$ with cost at most $c \cdot \cost(P, \opt_{k}(P \setminus D) \cup D)$.
Furthermore, we have:
    \begin{align*}
        \cost(P, \opt_{k}(P \setminus D) \cup D) &= \cost(P \setminus D, \opt_{k}(P \setminus D) \cup D)\\
        &\leq \cost(P \setminus D, \opt_{k}(P\setminus D)),
    \end{align*}
    where the equality holds as all points of $D$ contribute zero to $\cost(P, \opt_{k}(P \setminus D) \cup D)$, and the inequality because removing centers only increases the cost. Putting it all together, this shows 
     $\cost(P \setminus D, A) \leq c \cdot \cost(P \setminus D, \opt_{k}(P\setminus D))$.
\end{proof}

\section{$O(T/k)$ update time via merge-and-reduce tree}
We start by showing how \Cref{lem:singleCoreset} implies our main theorem \Cref{thm:main}.
For this, we sketch first the merge-and-reduce algorithm of \cite{Har-PeledM04}, and how to incorporate the coreset construction of \Cref{lem:singleCoreset} to speed up the update time.

\subsection{Description of the merge-and-reduce algorithm}
The goal of this algorithm is to maintain an $\eps$-coreset under insertions and deletions of points. The keys to this are the following strong properties of coreset: first, if $\coreset_1$ is an $\eps$-coreset for $P_1$, and $\coreset_2$ is an $\eps$-coreset for $P_2$, then $\coreset_1 \cup \coreset_2$ is an $\eps$-coreset for $P_1 \cup P_2$. Second, if instead $\coreset_2$ is an $\eps$-coreset for $\coreset_1$,  then $\coreset_2$ is a $(2\eps+\eps^2)$-coreset for $P_1$.

The merge-and-reduce data structure is the following (suppose for now that the number of points in the dataset stays within $[n, 2n]$). The dataset is partitioned into at most $2n/k$ parts, each containing at most $k$ points. Those parts form the leaves of a complete binary tree. We say that a node $v$ of the tree \emph{represents} the points stored at the leaves descendants of $v$.

Each node $v$ maintains a set $\coreset_v$ as follows. Let $v_1, v_2$ be the children of $v$. Node $v$ stores an $\eps$-coreset of $\coreset_{v_1} \cup \coreset_{v_2}$.
It follows from the two coreset properties that the set stored at the root is an $O(\log n \cdot \eps)$-coreset of the full dataset.\footnote{This can be shown by induction: a node at height $h$ stores an $O(h \eps)$-coreset of the points it represents.} Rescaling $\eps$ by $\log n$ therefore ensures that the root stores an $\eps$-coreset. 
It is straightforward to maintain this data structure under insertions: simply add the new point to a leaf that contains less than $k$ points, and update the sets stored at all its ancestors. For deletions, simply remove the point from the leaf it is stored in, and update all its ancestors.
Since the depth of the tree is $O(\log n)$, this triggers $\tilde O(1)$ coreset computations, every time on a dataset of size $2\coresetSize$ (where $\coresetSize$ is the size of an $\eps/\log n$-coreset, which is $\tilde O(k \eps^{-2})$). 
The update time is therefore the time to compute $\log(n)$ times an $\eps/ \log n$-coreset on a dataset of size $2\coresetSize$. This turns out to be $\tilde O(T(2\coresetSize, k))$ using standard coreset construction.
Finally, if the size of the dataset changes too much and jumps out of $[n, 2n]$, the algorithm recomputes from scratch the data structure with a fresh estimate on the number of points. \cite{Har-PeledM04} shows that the amortized complexity induced by this step stays $\tilde O(T(\coresetSize, k))$.

\subsection{Our algorithm}\label{sec:fullAlg}
The previous algorithm uses a static coreset construction as a black-box. Instead, we propose to use \Cref{lem:singleCoreset} to avoid reprocessing from scratch at every node. For each node, we divide time into \emph{epochs}, with the following property. For a node $v$ with children $v_1$ and $v_2$, the sets maintained at $v_1$ and $v_2$ change at most $k$ times during an epoch. When those sets have changed $k$ times, a  new epoch is started for the node and all its ancestors.

Each node maintains an $\eps$-coreset $\coreset$ of its two children using the algorithm from \Cref{lem:singleCoreset}. For each update of those, this algorithm processes them and transmits to its parent the potential updates made in $\coreset$. 
This is enough to show \Cref{thm:main}:

\begin{proof}[Proof of \Cref{thm:main}]
We let $n_i$ be the number of insertions to the dataset, and $n_d$ the number of deletions. Our goal is to show that the total running time is $\tilde O\lpar (n_i + n_d) \cdot \frac{T(k \polylog(n), k)}{k}\rpar$.

    At each node, the complexity can be decomposed into the one due to updates in between epochs, and the complexity due to starting new epochs. We say that the work done by a node is the total complexity to update its coreset, and the total complexity when starting a new epoch at this node (i.e., re-initializing this node and all its parents).

    We will compute the work done at each level of the tree: leaves have level $0$, and parent of a node at level $i$ has level $i+1$. For a node $v$, let $n_i^v, n_d^v$ be the total number of insertions and deletions in the dataset represented by $v$.
    
    We show the following claim by induction: the total work for a node $v$ at level $\ell$ is $\tilde O\lpar (n_i^v + n_d^v) \cdot \frac{T(k \polylog(n), k)}{k}\rpar$, the number of insertions to the coreset is $n_i^v + \ell \cdot n_d^v$, and the number of deletion is $n_d^v$. 

    For leaves (i.e., nodes at level $0$), the statement is straightforward, as the set they maintain is directly the dataset they represent, with weights 1. 

    Let $\ell \geq 1$, $v$ be a node at level $\ell$ with children $v_1, v_2$.  We let $n_i^{\coreset_v}, n_d^{\coreset_v}$ be the number of insertions and deletions made to the coresets maintained at $v_1$ and $v_2$. 
    By the induction hypothesis, the coresets $\coreset_1$ and $\coreset_2$ maintained at $v_1$ and $v_2$ have undergone $n_i^{\coreset_v} + (\ell-1) \cdot n_d^{\coreset_v}$ insertions, and $n_d^{\coreset_v}$ deletions.
    
    \Cref{lem:singleCoreset} therefore shows that, during each epoch, the total complexity to maintain a coreset at $v$ is $\tilde O\lpar (n_i^{\coreset_v} + n_d^{\coreset_v}) \cdot \frac{T(k \polylog(n), k)}{k}\rpar$, the number of insertions to the coreset at most $n_i^{\coreset_v} + n_d^{\coreset_v}$, and the number of deletions to the coreset at most $n_d^{\coreset_v}$.
    Therefore, over all epochs, the complexity is $O(n_i^v + \ell n_d^v)$, the number of insertions to the coreset is $n_i^v + \ell \cdot n_d^v$ and number of deletions at most $n_d^v$. 
    
    A new epoch is started at the node either when one of its descendant triggered a re-initialization, in which case the complexity is accounted at that descendant's level, or when $k$ updates have been made to the input during the current epoch.
    The latter occurs therefore at most  $(n_i^{\coreset_v} + \ell n_d^{\coreset_v})/k$ times, and every time triggers a recomputation of the coresets at $\log(n) - \ell$ many levels: therefore the complexity of a single recomputation is $\tilde O(T(\coresetSize, k, \eps/\log n) \cdot (\log n - \ell)) = \tilde O(T(k \polylog(n), k, \eps / \log n))$ (from the guarantee of \Cref{lem:singleCoreset}).
    Thus, the total complexity due to all re-initialization after $n_i^v$ insertions and $n_d^v$ deletions is $\tilde O \lpar T(k \polylog(n), k, \eps / \log n) \cdot  \frac{n_i^v + n_d^v}{k} \rpar$, which concludes the induction statement and the proof of the theorem.
\end{proof}

\section{An efficient dynamic coreset algorithm}

In this section, we show the key \Cref{lem:singleCoreset}. 
For this,  we describe an algorithm to maintain a coreset of size $k \poly(\log n, \eps^{-1})$ for a set of $O(n)$ points with weights in $[1, 2]$, in amortized time $\frac{T(n, k, \eps)}{k}$. 
As explained in the previous section, this algorithm can be used black-box to reduce the complexity to $\frac{T(\tilde O(k), k, \eps)}{k}$.

\subsection{The Algorithm}
We explain first the data structure that is maintained by the algorithm, and how to extract efficiently a coreset from it. We will then show how to initialize the data structure, and maintain it under insertions and deletions.

\paragraph*{The data structure.}

Let $P_0$ be the initial set of points, $P$ the current set of points, $I$ (resp.~$D$) the set of points inserted (resp.~deleted) since the beginning. For the lemma, we focus on the case where $|I| + |D| \le k$.

The data structure consists of a set of centers $A$, a scalar $\Delta$, and a set of groups $G_{small}, G_{close}, G_1, G_2, ...$ with the following guarantees:
\begin{enumerate}
    \item\label{prop:solApprox} $A$ consists of $O(k)$ centers such that $\cost(P_0 \setminus D, A) = O(\opt_k(P \setminus D))$, and $\Delta \in \R^+$ verifies $\Delta \leq \frac{\cost(P_0 \setminus D, A)}{|P|}$,
    \item\label{prop:maintainGroups}  $G_{small}, G_{close}, G_1, G_2, ...$  partition $P_0 \setminus D$ such that 
    \begin{itemize}
        \item $|G_{small}|\leq k \poly(\log n, \eps^{-1})$,
        \item $\forall p \in G_{close}, \cost(p, A) \leq \eps \Delta$, and 
        \item for all $i \geq 1$, $G_i$ and $A$ satisfy \Cref{prop:groups}.
    \end{itemize}
    \item\label{prop:random} For each group its points are maintained in random order in a data structure that allows for efficient insertions and deletions.
    \item\label{prop:weight} Finally, for each group $G_i$  with $i = 1, ...$ a lazy estimate $c_i$ on $|G_i|$ is maintained.
\end{enumerate}

Given this data structure, we can easily build a coreset of size $k \poly(\log n, \eps^{-1})$, as we sketch here and prove in \Cref{lem:oneLevelCoreset}. $G_{small}$ is a coreset for itself, and $A$ is a coreset for $G_{close}$. For each other group $G_i$ for $i \geq 1$,
\Cref{lem:coresetStatic} shows that the first $\coresetSize$ points of the random order, with weights multiplied by $|G_i| / \coresetSize$, form a coreset for $G_i$. 
Since the groups partition $P \setminus D$, the union of those coresets is a coreset for $P\setminus D$. 
To get a coreset for $P$, one merely needs to add each point of $I$ with weight $1$.

However, if we used the weight $w(p) |G_i|/ \coresetSize$ for each point $p$ in $G_i$, the weights of a coreset point, and thus the coreset itself, would change too frequently. Thus, instead we use $w(p) c_i / \coresetSize$ as weight for each coreset point.
We describe below how the data structure maintains lazily an estimate $c_i$ of the size of each $G_i$, while still guaranteeing that $c_i$ is a good estimate of $|G_i|$.

We first show how our algorithm maintains this data structure, and prove in the next section that this indeed maintains efficiently a coreset.  

\paragraph*{Initialization at the beginning of an epoch.}
For initialization, the algorithm computes an $O(1)$-approximation $A$ to $2k$-means on $P_0$.  Define $E_0$ as the $k$ most expensive points of $P_0$ in this solution $A$. $\Delta$ is set to be the average cost of $P_0 \setminus E_0$, i.e., $\cost(P_0 \setminus E_0)/|P_0 \setminus E_0|$. 

Next the algorithm defines groups as follows:\footnote{For convenience, we index the groups by three integers $j, b, w$ instead of a single one as in the previous description.}
\begin{itemize}
\item For $i = 1, ..., 2k$, let $a_i$ be the $i$-th center of $A$, and define $C_i$ to be the cluster of $a_i$, namely all points of $P_0 \setminus E_0$ closer to $a_i$ than to any other center of $A$ (breaking ties arbitrarily).
\item For $j \geq 1$, let $R_{i,j} := \{ x\in C_i : 2^{j-1} \cdot \eps \sqrt{\Delta} \leq \|x-a_i\|^2 < 2^j \cdot \eps \sqrt{\Delta} \}$. Let also $R_{i, 0} := \{ x\in C_i : \|x-a_i\| \leq \eps \sqrt{\Delta}\}$.
Each $R_{i,j}$ for $i = 1, ... , 2k$ is called a \emph{$j$-th ring}.
Note that the $R_{i,j}$ partition $P_0 \setminus E_0$.
\item For $w \geq 1$, let $R_{i,j,w} = \{p \in R_{i,j} : w(p) \in [2^w, 2^{w+1})\}$.
\item For $b \geq 0$, and all $w, j \geq 1$, let $G_{j, b, w} := \{ x: \exists i, x \in R_{i, j, w}, 2^b \leq |R_{i,j,w}| < 2^{b+1}\} = \cup_{i, 2^b \le |R_{i,j,w}| < 2^{b+1}} R_{i,j,w}$ be the set of $j$-th rings with cardinality between $2^b$ and $2^{b+1}$. $c_{j,b,w}$ is set to be $|R_{i,j,w}|$.
\item We say a group $G_{j,b,w}$ is \textit{initially large} when its size at the beginning of the epoch is more than $\log(n) \coresetSize / \eps$, otherwise the group is \textit{initially small}.
The group $G_{small}$ consists of the points of all  initially small groups, together with $E_0$. 
\item The group $G_{close}$ is initialized to be $\cup_i R_{i, 0}$. 
\end{itemize}
Finally, in each of the initially large groups, the algorithm randomly orders points and store them in a binary search tree (to allow for efficient insertions and deletions). Each point is associated with a random number in $[0, 1]$, and those are stored in a binary search tree with keys being the random number. 

We make a few remarks about this construction:
\begin{fact}\label{fact:init}
(1) The groups form a partition of $P_0$. (2) There are only $\log^3 (n/\eps)$ non-empty groups. 
(3) Each $G_{j,b,w}$ and $A$ satisfy \Cref{prop:groups}  at the beginning of the epoch
    As long as $|D|\leq k$, it holds that (4)  $\cost(P_0 \setminus D, A) = O(\opt_k(P \setminus D))$, and (5) $\Delta$ is smaller than the average cost of $P_0 \setminus D$.
\end{fact}
\begin{proof}
    (1) and (3) are direct consequences of the definition of the groups. 
    For (2), note that for any $i$ and $j \geq \log(|P_0 \setminus E_0|/\eps)$, the set $R_{i,j}$ is empty (as the cost of points in such a ring would be larger than $\cost(P_0\setminus E_0)$, which is a contradiction). Furthermore, no ring can contain more than $|P_0 \setminus E_0|$ points. Lastly, the lemma's condition ensures that all weights are in $[1, 2n]$. Therefore, groups with $b > \log(n)$, or $j > \log(n/\eps)$, or $w > \log(2n)$ are empty: this concludes the second bullet.

    Finally, (4) holds directly from \Cref{lem:bicriteria}, and (5) stems from the definition of $\Delta$ and the choice of $E_0$.
\end{proof}

Therefore, the groups form a partition of $P_0$. 
Note that there are there are $\log^2 n$ many groups $G_{j,b,w}$ and, thus, 
$G_{small}$ contains at most $\log^3(n) n_C/\epsilon$ many points.

\paragraph*{Processing insertions.}
Dealing with insertion is easy: the set of inserted points does not appear in the definition of the data structure.
Therefore, as long as the total number of updates is less than $k$, the data structure needs no update after an insertion. When $|I|+|D| > k$, the algorithm is done (and, in the use of this subroutine in \Cref{sec:fullAlg}, a new ``epoch" starts).

\paragraph*{Processing deletions.}
Property \ref{prop:solApprox} is a consequence of \Cref{fact:init}.
Properties \ref{prop:maintainGroups}  and \ref{prop:random} of the data structure are more involved, as we need to ensure that the groups still fulfill \Cref{prop:groups}, e.g., their size stays roughly the same. 
The algorithm updates the groups as follows.
First, if the deleted point is in a group that is initially small, then no further updates are required.

If the point is in a group that is initially large, the algorithm only needs to ensure that all clusters in a group have the same size, up to a factor $8$.  
Suppose the deleted point was in a ring $R_{i,j,w}$ from the group $G_{j, b,w}$. Note that two things can happen. Either the number of points in the ring after deletion is still more than $2^{b-3}$, in which case \Cref{prop:groups}  still holds and the algorithm does nothing. 
If, however, the number of points becomes exactly $2^{b-3}$,
the algorithm removes the whole ring from $G_{j,b,w}$ and adds it to $G_{j, b-3, w}$ (note that, in that case, $b-3 \geq 0$ and $G_{j,b-3, w}$ exists).

This movement triggers some necessary changes in the ordering of each group, to maintain Property \ref{prop:random} of the data structure, i.e., the random order of the points of each group in the data structure. 
First, in the group $G_{j, b, w}$, the points of the ring are simply removed. 
Second, they are inserted one by one at positions that are chosen uniformly at random into $G_{j,b-3,w}$. This can be done efficiently, as the orderings are described with a binary search tree.

Finally, the size estimate of the two groups may have to be updated: if $|G_{j,b,w}| \leq (1-\eps) c_{j,b,w}$ set $c_{j,b,w} \gets |G_{j,b,w}|$. 
If a ring was moved in the group $G_{j, b-3, w}$ and if $(1+\eps) c_{j,b-3,w} \leq |G_{j,b-3,w}|$, then set $c_{j,b-3,w} \gets |G_{j,b-3,w}|$.
Both cases imply that the weights of all coreset points from the group considered change.

\paragraph*{Extracting a coreset from the data structure}
From the data structure described above, the algorithm extracts a coreset as follows. First, it defines weights $w$ such that:
\begin{itemize}
    \item  $w(p) = 1$ for $p \in I$
    \item  $w(p) = 1$  for $p \in G_{small}  \setminus D$
    \item for all centers $a_i$ of $A$, $w(a_i) = |R_{i, 0} \setminus D|$,
    \item for each point $p$ in an initially large group $G_{j,b,w}$, $w(p) = c_{j,b,w}$ if $p$ is among the first $\coresetSize$ elements of $G_{j,b,w}$ in the random order of the data structure, and 0 otherwise,
    \item $w(p) = 0$ for each other point $p$.
\end{itemize}
Let $\coreset$ be the set of points with non-zero weight. Those weights can be easily maintained under insertions and deletions, as described in the previous paragraphs.
We first show that $\coreset$ can be computed in amortized $\tilde O(1)$ time, then that it is an $\eps$-coreset for $P_0 \setminus D \cup I$.

\subsection{Running-time Analysis}

To show the running time, we assume for simplicity of the bounds $\eps \leq 1/3$, and $\coresetSize \geq 10 k \eps^{-2}$. The second assumption is (almost) without loss of generality, as in general metric spaces any $\eps$-coreset must have size $\Omega(k \eps^{-2} \log n)$ \cite{Cohen-AddadLSS22}, and in Euclidean space they must have size $\Omega(k \eps^{-4})$~\cite{huang23optimal}.

We start by listing the running time induced by each of the operations described in the algorithm. 
The initialization takes time $T(n, 2k)$ to compute the constant-factor approximation to $2k$-means, and assignment of each point to its closest center. Then, the identification of $E_0$ and partitioning into groups take linear time. This is enough to prove the first statement of \Cref{lem:singleCoreset}

Any insertion just requires to set the weight of the point to $1$, which takes constant time. Deletions are more intricate to analyze.
We first note that the running time to insert or remove a point from a group of the data structure takes time $\tilde O(1)$, which is the time to remove the point from a binary search tree and add it to another one.

It may happen that a ring $R_{i,j,w}$ is moved from one initially large group $G_{j,b,w}$  to $G_{j,b-3,w}$.
In that case, the running time is $\tilde O(|R_{i,j,w}|)$, in order to move all points of $R_{i,j,w}$;
and the coreset changes by at most $2|R_{i,j,w}|$: first, at most $|R_{i,j,w}|$ points are removed from the coreset of $G_{j,b,w}$ (and replaced by points that stay in $G_{j,b,w}$ after the operation). Second, the $|R_{i,j,w}|$ points that are added to $G_{j,b,w}$ are placed randomly in the random order: if they appear among the first $n_c$ elements, they need to be added to the coreset of $G_{j,b-3,w}$. Therefore, there are at most $|R_{i,j,w}|$ many changes in the coreset of $G_{j, b-3, w}$ (the same argument works if $G_{j,b-3,w}$ is initially small). Therefore, in total the coreset changes by at most $2|R_{i,j,w}|$.

Finally, the size estimate for a group $G_{j,b,w}$ may appear to change. However, as we will demonstrate in \Cref{lem:sizeDDecrease}, it is never actually decreased.
The size estimate increases when several rings have been moved to the group: in that case, the weight of up to $\coresetSize$ many coreset points changes.
In the amortized analysis below, we will show that those costly events do not occur too often, which will conclude the proof of \Cref{lem:oneLevelCoreset}.

{\bf  Amortized analysis.}
To analyze the number of changes to the coreset by this algorithm, we proceed with a token-based argument. Every deletion of a point from the dataset 
is charged one token, i.e., it increases the number of tokens by one.
The tokens are used to bound the total number of updates in the coreset: whenever a point is removed from the coreset, it will consume one token. This can happen both for its deletion, or when its weight is updated (which we see as deleting the point and re-inserting it with a different weight).

To proceed with the analysis, we define \emph{token wallets} of several types. 
For each group, there is one wallet $T^G_{j,b,w}$, used to update the weight of the group. For each ring $R_{i,j,w}$, there is a wallet $T^R_{i,j,w}$ which is used for deletions occurring when moving a ring to another group, and $T^i_{i,j,w}$ which is used as an intermediary wallet to supply $T^G_{j,b,w}$.

Tokens are provided to the wallets when points are deleted from the dataset: each deleted point $p$ brings one token as follows.
Let $R_{i,j,w}$ be the ring of $p$.
If $p$ has non-zero weight, then it directly consumes its token to pay for its own deletion from the coreset. Otherwise, it gives $1/2$ token to $T^R_{i,j,w}$, and $1/2$ to $T^i_{i,j,w}$.
When a ring $R_{i,j,w}$ moves to a group $G_{j,b,w}$, all tokens of $T^i_{i,j,w}$ are transferred to $T^G_{j,b,w}$. 

To show that those tokens are enough to pay for deletions, we use a probabilistic analysis using the randomness of our algorithm: in each group, points are sorted randomly, which will ensure that deletions from the input rarely triggers deletions in the coreset, as we show in the next lemma.

\begin{lemma}\label{lem:tokenRing}
    Let $b \in [3, \dots, \log n]$, $i \in [1,k]$, and $j \in [1,\log n]$.
    Consider a ring $R_{i,j,w}$ in group $G_{j,b,w}$ that is initially large, and let $m = 2^{b} - 2^{b-3}$. Let $p_1, \dots, p_m$ be the first $m$ points removed from the ring $R_{i,j,w}$. Then, after those $m$ deletions, both $T^R_{i,j,w}$ and $T^i_{i,j,w}$ contain at least $2^{b-3}$ tokens with probability at least $1-1/n^{6}$.
\end{lemma}
\begin{proof}
    First, the lemma statement is well defined: when $R_{i,j,w}$ was placed into $G_{j,b,w}$ (either at the beginning of the epoch, or after points were removed from $R_{i,j,w}$), its size was at least $2^b$, and therefore there are indeed at least $m$ points in the ring. 

    Each of the deleted points contributes $1/2$ to $T^R_{i,j,w}$ and $T^G_{j,b,w}$, if it is not part of the coreset. Therefore, the number of tokens added by those points in both wallets is the same, and it is enough to analyze $T^R_{i,j,w}$.  
    Let $X_i$ be the random variable equal to $1/2$ if $p_i$ is not part of the coreset when it is deleted: we have $|T^R_{i,j,w}| = \sum_{i=1}^m X_i$. We will show that this sum of variables is a martingale, and use Azuma's inequality to conclude.

    We make a few observations. Since the group is initially large, its size is at least $\log n  \cdot \coresetSize / \eps$.
    All rings in the group have initially size between $2^{b}$ and $2^{b+1}$. As there are initially at most $k$ rings in the  group $G_{b,j,w}$ this ensures $2^b \geq \frac{\log n  \cdot \coresetSize}{k \eps} \geq \frac{\log(n)}{\eps^3}$. 
    Using $\eps \leq 1/3$, this ensures $m = 2^b - 2^{b-3} = (7/8)2^b \geq 230\log(n)$.
    
    It seems more natural to consider the process of deletions  reversed. Starting from the set of points $G_{j,b,w} \setminus \lbra p_1, \dots p_m \rbra$, points are added sequentially at random locations, in order $p_m, p_{m-1}, ..., p_1$. 
    Then, $X_i = 1/2$ when point $p_i$ is not inserted among the first $\coresetSize$ locations.
    This is equivalent to the initial process as the relative positions of $p_i, p_{i+1}, ...$ in the random order do not depend on the positions of $p_1, ..., p_{i-1}$.
    Furthermore, this shows $X_i$ is independent of $X_{i+1}, ..., X_m$: the position where $p_i$ is inserted does not depend on the relative order of $p_{i+1}, \dots, p_m$. 

    We use this fact to show that the number of tokes follows a martingale. Formally, let $\mu_i = E[\sum_{j=i}^m X_j]$, let  $Y_i = \sum_{j=i}^m X_j - \mu_i$ for $1 \le i \le m$, and let $Y_{m+1} = 0$. Then $E[Y_i] = 0$  and $E[Y_i | Y_{i+1}, \dots , Y_m] =
    E[X_i -E[X_i]| Y_{i+1}, \dots , Y_m] + E[Y_{i+1} |Y_{i+1}, \dots , Y_m] = E[X_i -E[X_i]] + Y_{i+1}=
    Y_{i+1}$.
    Therefore, the sequence of variables $Y_i$ for $i = m, m-1, ..., 1$ is a martingale. We will  use concentration bound on martingales to prove the lemma.
    
    Note that for each $i \in [1,m],~X_i=0$ with probability at most $\eps$: indeed, when $p_i$ is added, there are already at least $|G_{j,b,w}|-m = 2^{b-3} \geq \coresetSize / \eps$  points in the order (using $\log(n) \ge 8$ and that $G_{j,b,w}$ is initially large), and therefore $\eps$ is an upper-bound on the probability of $p_i$ being added among the first  $\coresetSize$ positions. 
    Thus, for each $i \in [1,m]$, $1/2 \ge E[X_i] \ge (1-\epsilon)/2$ and we have $\mu_1 = \E[\sum_{j=1}^{m+1} X_j] \geq (1-\eps)m/2$.
    
     As $|Y_i - Y_{i+1}| = |X_i -E[X_i]| \le 1/2$ for each $i \in [1,m]$, 
    Azuma's inequality ensures that, with probability at most $\exp(-2t^2/m)$, $Y_1 - Y_{m+1} \le -t$, which is equivalent to $\sum_{j=1}^m X_j \leq \mu_1 - t$. 
    It follows that $\sum_{j=1}^m X_j \leq (1-\eps) m/2 - t$, with probability at most $\exp(-2t^2/m)$.

    Taking $t=m/8$ and using $\eps \leq 1/3$ yields that, with probability at least $1-\exp(-m/32)$,  $\sum_{j=1}^m X_j \geq (2/3)m/2 -m/8 \geq m/5 = (2^b-2^{b-3})/5 \geq 2^{b-3}$. 
    As $m \geq 230\log n$, this probability is at least 
    $1-1/n^6$.
    As the number of tokens is exactly $\sum_{j=1}^m X_j$, this concludes the proof.
\end{proof}

As a corollary of the previous lemma, we can compute the number of tokens in $T^G_{j,b,w}$:

\begin{corollary}\label{lem:tokenGroups}
    Consider a group $G_{j,b,w}$, at the moment its size estimate increases. Then, $T^G_{j,b,w}$ contains at least $\coresetSize$ tokens with probability at least $1-1/n^5$.
\end{corollary}
\begin{proof}
    Elements can be added to a group $G_{j,b,w}$ only by ring movements: rings can move from $G_{j, b+3,w}$ to $G_{j,b,w}$ when they contain $2^{b}$ elements. \Cref{lem:tokenRing} ensures that, for every such ring movement, $2^{b}$ tokens are added  to $T^G_{j,b,w}$ with probability at least $1-1/n^6$. 
    Since a ring movement adds exactly $2^{b}$ points to $G_{j,b,w}$, there is exactly one token per new point. 
    Furthermore, during an epoch there can be at most $k$ deletions, and therefore at most $k$ ring movements.
    A union-bound ensures that the previous holds for all ring movement with probability $1-k/n^6 \geq 1-1/n^5$. 

    When the estimated weight of $G_{j,b,w}$ changes, it means that its size increased by an $\eps$ fraction: since the group is initially large, it means at least $\log(n) \cdot \coresetSize$ elements have been added to it, and therefore $T^G_{j,b,w}$ contains at least $\log(n) \cdot \coresetSize$ many tokens. This concludes the corollary.
\end{proof}

In addition to this token scheme, we have the following property:
\begin{lemma}\label{lem:sizeDDecrease}
    For any group $G_{j,b,w}$ that is initially large, the size estimate $c_{j,b,w}$ does not decrease during a sequence of at most $k$ updates.
\end{lemma}
\begin{proof}
    If $G_{j,b,w}$ is initially large, it contains at least $\log(n) \cdot \coresetSize /\eps$ many points. 
    Points can be removed from the group for two reasons: either they are deleted from the dataset, or their ring moves to another group. Let $t_1$ be the number of points deleted from the dataset, and $t_2$ the number of points removed because their ring moved. Note that when a ring moves, at least $3/4$ of its point have been deleted from the dataset.  Therefore, if the size of the group decreases by some number $t$, at least $t_1 + 3t_2/4 \geq 3t/4$  points have been deleted from the dataset. Since we know that at most $k$ elements are deleted, this enforces $t \geq 4k/3$.

    Now, the algorithm decreases $c_{j,b,w}$ if the group sizes decreases by an $\eps$ factor, which means $t \geq \log(n) \cdot \coresetSize = \log(n) k / \eps^4$ points have been deleted from the group. The previous paragraph shows that this cannot happen with only $k$ deletions from the input.
\end{proof}

We can now proceed to the proof of \Cref{lem:singleCoreset}.
\begin{proof}[Proof of \Cref{lem:singleCoreset}]
    As presented in the initial sketch, the running time for initializing the data structure is $\tilde O(T(n, 2k))$, and each insertion is processed in constant time and yields a single addition to the coreset.

    The effect of deletions can be bounded via the previous lemmas as follows. 
    First, when the point deleted by the update is part of the coreset, it is removed from the data structure in time $\tilde O(1)$, and uses one token to account for the update in the coreset.
    
    When a ring moves from $G_{j,b,w}$ to $G_{j,b-3,w}$, its size is exactly $2^{b-3}$: the complexity of processing the movement (removing points of the ring from $G_{j,b,w}$ and adding them to $G_{j, b-3,w}$) is therefore $\tilde O(2^b) = \tilde O(|T^R_{i,j,w}|)$. The tokens in $T^R_{i,j,w}$ are therefore enough to pay for the running time.  Similarly, the number of insertions and deletions in the coreset is at most $2 \cdot 2^{b-3} \leq |T^R_{i,j,w}|$.

    When the size estimate of $G_{j,b,w}$ changes, the weights 
    of $\coresetSize$ many points are updated: this has running time $O(\coresetSize)$. 
    \Cref{lem:sizeDDecrease} shows that the size must have increased, and \Cref{lem:tokenGroups} ensures that $T^G_{j,b,w} \geq \coresetSize$. Those tokens can therefore pay for the updates.

    Since the total number of tokens is the number of deleted points, this concludes the lemma.
\end{proof}

\subsection{Correctness analysis}
\begin{lemma}\label{lem:oneLevelCoreset}
    For group $G_{j,b,w}$, let $\coreset_{j,b,w}$ be the $\coresetSize$ first points of the random order maintained by the data structure. Then, $E_0 \setminus D \cup I  \bigcup_{j,b,w} \coreset_{j,b,w}$ with weights defined in the algorithm is a $2\eps + \eps^2$-coreset for $P_0 \setminus D \cup I$.
\end{lemma}
\begin{proof}
    We use the composability property of coreset. $I$ with weights $1$ is obviously a coreset for $I$. Similarly, $E_0  \setminus D$ is a coreset for $E_0 \setminus D$, and $G_{small}$ a coreset for $G_{small}$.

    Let $\Delta$ be as in the algorithm, the average cost of $P_0 \setminus E_0$.
    The algorithm ensures that every point in $G_{close}$ is at distance at most $\eps \Delta$ of a center of $A$.
    Item (2) of \Cref{fact:init} shows that $\Delta \leq \cost(P_0 \setminus D) / |P_0 \setminus D|$: therefore, the triangle inequality ensures that each point of $G_{close}$ can be replaced by its closest center in $A$, up to a total error $\eps |G_{close}| \Delta \leq \eps \cdot \cost(P_0\setminus D, A) \leq O(\eps) \opt_k(P_0 \setminus D)$. The centers of $A$, weighted by $|R_{i, 0} \setminus D|$, form therefore an $\eps$-coreset for $G_{close}$. 
    
    The first item of \Cref{fact:init} ensures that $A$ is an $O(1)$-approximation to $k$-means on $P_0 \setminus D$: therefore, \Cref{lem:coresetStatic} shows that for each initially large group $G_{j,b,w}$,  $\coreset_{j,b,w}$ with weight per point $w_{j,b,w} := \frac{|G_{j,b,w}|}{\coresetSize}$ 
    is an $\eps$-coreset for $G_ {j,b,w}$.
    Since the total weight estimates satisfy $c_{j,b,w} = (1\pm \eps) |G_{j,b,w}|$, each $\coreset_{j,b,w}$ with weights $w_{j,b,w} := \frac{c_{j,b,w}}{\coresetSize}$ is a $2\eps+\eps^2$-coreset for $G_{j,b,w}$.
    
    The union of all those coresets is therefore a coreset for $P_0 \setminus D \cup I$.
\end{proof}

\section{A Note on Euclidean Spaces}
\label{sec:euclideanSpace}
In Euclidean space, it is known how to compute an $O(\polylog(k))$-approximation to $(k,z)$-clustering, in time essentially $\tilde O(nd)$, ignoring the impact of the aspect-ratio for our discussion (see Corollary 4.3 in \cite{Cohen-AddadLNSS20} for the running time, combined with Lemma 3.1 for the approximation ratio, with the dimension reduced to $O(\log k)$ using \cite{MakarychevMR19}).
To get an $O(1)$-approximation, \cite{LattanziS19, ChooGPR20} showed how to complement the famous $k$-means++ algorithm with few local search steps, achieving a running time $\tilde O(nd + nk)$. 

Therefore, we believe that the following conjecture holds: 
\begin{conjecture}
    There is an $O(1)$-approximation algorithm for $(k,z)$-clustering in Euclidean space $\R^d$, running in time $\tilde O(nd)$. 
\end{conjecture}
In that case, the query time of our fully dynamic algorithm reduces to $\tilde O(k)$, i.e., time to output the solution.

\section{Conclusion}
We present an efficient algorithm to maintain a coreset under insertion and deletion of points. This algorithm has near-optimal running time, as it can be used black-box to solve $(k,z)$-clustering with optimal update time (and improving ours would directly improve the update time for $(k,z)$-clustering).

Furthermore, our algorithm likely yields an optimal algorithm for update and query time in Euclidean space. This is true, under two conjectures  that we believe are worth investigating.: first, there exists a static $O(1)$-approximation algorithm for $(k,z)$-clustering with running time $\tilde O(nd)$ (which we discuss in \Cref{sec:euclideanSpace}). Second, the query time is at least $\Omega(k)$. 
This is the time to output a solution; however, it may be the case that solutions do not entirely change between each query.

This second conjecture is thus related to the notion of \emph{consistency}: \cite{FichtenbergerLN21} recently showed how to maintain a $k$-median solution in an insertion-only stream of length $n$, with at most $\tilde O(k)$ total number of changes in the solution. 
It is therefore natural to ask whether this approach can be extended to fully-dynamic streams: is it possible to maintain a solution with at most $O(1)$ changes between each time step?

\section*{Acknowledgments}
\erclogowrapped{5\baselineskip}Monika Henzinger:  This project has received funding from the European Research Council (ERC) under the European Union's Horizon 2020 research and innovation programme (Grant agreement No. 101019564) and the Austrian Science Fund (FWF) grant DOI 10.55776/Z422, grant DOI 10.55776/I5982, and grant DOI 10.55776/P33775 with additional funding from the netidee SCIENCE Stiftung, 2020–2024.

This work was partially done while David Saulpic was at the Institute for Science and Technology, Austria (ISTA). David Sauplic has received funding from the European Union’s Horizon 2020 research and innovation programme under the Marie Sklodowska-Curie grant agreement No 101034413. 
This work was partially funded by the grant ANR-19-CE48-0016 from the French National Research Agency (ANR).  

\bibliography{biblio}

\begin{thebibliography}{10}

\bibitem{agarwal2004approximating}
Pankaj~K Agarwal, Sariel Har-Peled, and Kasturi~R Varadarajan.
\newblock Approximating extent measures of points.
\newblock {\em Journal of the ACM (JACM)}, 51(4):606--635, 2004.

\bibitem{BateniEFHJMW23}
MohammadHossein Bateni, Hossein Esfandiari, Hendrik Fichtenberger, Monika
  Henzinger, Rajesh Jayaram, Vahab Mirrokni, and Andreas Wiese.
\newblock Optimal fully dynamic $k$-center clustering for adaptive and
  oblivious adversaries.
\newblock In Nikhil Bansal and Viswanath Nagarajan, editors, {\em Proceedings
  of the 2023 {ACM-SIAM} Symposium on Discrete Algorithms, {SODA} 2023,
  Florence, Italy, January 22-25, 2023}, pages 2677--2727. {SIAM}, 2023.
\newblock URL: \url{https://doi.org/10.1137/1.9781611977554.ch101}, \href
  {https://doi.org/10.1137/1.9781611977554.CH101}
  {\path{doi:10.1137/1.9781611977554.CH101}}.

\bibitem{BentleyS80}
Jon~Louis Bentley and James~B. Saxe.
\newblock Decomposable searching problems {I:} static-to-dynamic
  transformation.
\newblock {\em J. Algorithms}, 1(4):301--358, 1980.
\newblock \href {https://doi.org/10.1016/0196-6774(80)90015-2}
  {\path{doi:10.1016/0196-6774(80)90015-2}}.

\bibitem{sayan23}
Sayan Bhattacharya, Mart{\'{\i}}n Costa, Silvio Lattanzi, and Nikos Parotsidis.
\newblock Fully dynamic k-clustering in $\tilde o(k)$ update time.
\newblock {\em To appear at NeurIPS"23}, 2023.
\newblock URL: \url{https://doi.org/10.48550/arXiv.2310.17420}.

\bibitem{BravermanFLSY17}
Vladimir Braverman, Gereon Frahling, Harry Lang, Christian Sohler, and Lin~F.
  Yang.
\newblock Clustering high dimensional dynamic data streams.
\newblock In Doina Precup and Yee~Whye Teh, editors, {\em Proceedings of the
  34th International Conference on Machine Learning, {ICML} 2017, Sydney, NSW,
  Australia, 6-11 August 2017}, volume~70 of {\em Proceedings of Machine
  Learning Research}, pages 576--585. {PMLR}, 2017.
\newblock URL: \url{http://proceedings.mlr.press/v70/braverman17a.html}.

\bibitem{BravermanJKW21}
Vladimir Braverman, Shaofeng~H.{-}C. Jiang, Robert Krauthgamer, and Xuan Wu.
\newblock Coresets for clustering in excluded-minor graphs and beyond.
\newblock In D{\'{a}}niel Marx, editor, {\em Proceedings of the 2021 {ACM-SIAM}
  Symposium on Discrete Algorithms, {SODA} 2021, Virtual Conference, January 10
  - 13, 2021}, pages 2679--2696. {SIAM}, 2021.
\newblock \href {https://doi.org/10.1137/1.9781611976465.159}
  {\path{doi:10.1137/1.9781611976465.159}}.

\bibitem{ChanGS18}
T.{-}H.~Hubert Chan, Arnaud Guerquin, and Mauro Sozio.
\newblock Fully dynamic $k$-center clustering.
\newblock In Pierre{-}Antoine Champin, Fabien Gandon, Mounia Lalmas, and
  Panagiotis~G. Ipeirotis, editors, {\em Proceedings of the 2018 World Wide Web
  Conference on World Wide Web, {WWW} 2018, Lyon, France, April 23-27, 2018},
  pages 579--587. {ACM}, 2018.
\newblock \href {https://doi.org/10.1145/3178876.3186124}
  {\path{doi:10.1145/3178876.3186124}}.

\bibitem{ChooGPR20}
Davin Choo, Christoph Grunau, Julian Portmann, and V{\'{a}}clav Rozhon.
\newblock k-means++: few more steps yield constant approximation.
\newblock In {\em Proceedings of the 37th International Conference on Machine
  Learning, {ICML} 2020, 13-18 July 2020, Virtual Event}, volume 119 of {\em
  Proceedings of Machine Learning Research}, pages 1909--1917. {PMLR}, 2020.
\newblock URL: \url{http://proceedings.mlr.press/v119/choo20a.html}.

\bibitem{Cohen-AddadHPSS19}
Vincent Cohen{-}Addad, Niklas Hjuler, Nikos Parotsidis, David Saulpic, and
  Chris Schwiegelshohn.
\newblock Fully dynamic consistent facility location.
\newblock In Hanna~M. Wallach, Hugo Larochelle, Alina Beygelzimer, Florence
  d'Alch{\'{e}}{-}Buc, Emily~B. Fox, and Roman Garnett, editors, {\em Advances
  in Neural Information Processing Systems 32: Annual Conference on Neural
  Information Processing Systems 2019, NeurIPS 2019, December 8-14, 2019,
  Vancouver, BC, Canada}, pages 3250--3260, 2019.
\newblock URL:
  \url{https://proceedings.neurips.cc/paper/2019/hash/fface8385abbf94b4593a0ed53a0c70f-Abstract.html}.

\bibitem{Cohen-AddadLSS22}
Vincent Cohen{-}Addad, Kasper~Green Larsen, David Saulpic, and Chris
  Schwiegelshohn.
\newblock Towards optimal lower bounds for k-median and k-means coresets.
\newblock In Stefano Leonardi and Anupam Gupta, editors, {\em {STOC} '22: 54th
  Annual {ACM} {SIGACT} Symposium on Theory of Computing, Rome, Italy, June 20
  - 24, 2022}, pages 1038--1051. {ACM}, 2022.
\newblock \href {https://doi.org/10.1145/3519935.3519946}
  {\path{doi:10.1145/3519935.3519946}}.

\bibitem{Cohen-AddadLSSS22}
Vincent Cohen{-}Addad, Kasper~Green Larsen, David Saulpic, Chris
  Schwiegelshohn, and Omar~Ali Sheikh{-}Omar.
\newblock Improved coresets for euclidean k-means.
\newblock In Sanmi Koyejo, S.~Mohamed, A.~Agarwal, Danielle Belgrave, K.~Cho,
  and A.~Oh, editors, {\em Advances in Neural Information Processing Systems
  35: Annual Conference on Neural Information Processing Systems 2022, NeurIPS
  2022, New Orleans, LA, USA, November 28 - December 9, 2022}, 2022.
\newblock URL:
  \url{http://papers.nips.cc/paper\_files/paper/2022/hash/120c9ab5c58ba0fa9dd3a22ace1de245-Abstract-Conference.html}.

\bibitem{Cohen-AddadLNSS20}
Vincent Cohen{-}Addad, Silvio Lattanzi, Ashkan Norouzi{-}Fard, Christian
  Sohler, and Ola Svensson.
\newblock Fast and accurate $k$-means++ via rejection sampling.
\newblock In Hugo Larochelle, Marc'Aurelio Ranzato, Raia Hadsell,
  Maria{-}Florina Balcan, and Hsuan{-}Tien Lin, editors, {\em Advances in
  Neural Information Processing Systems 33: Annual Conference on Neural
  Information Processing Systems 2020, NeurIPS 2020, December 6-12, 2020,
  virtual}, 2020.
\newblock URL:
  \url{https://proceedings.neurips.cc/paper/2020/hash/babcff88f8be8c4795bd6f0f8cccca61-Abstract.html}.

\bibitem{stoc21}
Vincent Cohen{-}Addad, David Saulpic, and Chris Schwiegelshohn.
\newblock A new coreset framework for clustering.
\newblock In Samir Khuller and Virginia~Vassilevska Williams, editors, {\em
  {STOC} '21: 53rd Annual {ACM} {SIGACT} Symposium on Theory of Computing,
  Virtual Event, Italy, June 21-25, 2021}, pages 169--182. {ACM}, 2021.
\newblock \href {https://doi.org/10.1145/3406325.3451022}
  {\path{doi:10.1145/3406325.3451022}}.

\bibitem{Cohen-AddadWZ23}
Vincent Cohen{-}Addad, David~P. Woodruff, and Samson Zhou.
\newblock Streaming euclidean k-median and k-means with o(log n) space.
\newblock In {\em 64th {IEEE} Annual Symposium on Foundations of Computer
  Science, {FOCS} 2023, Santa Cruz, CA, USA, November 6-9, 2023}, pages
  883--908. {IEEE}, 2023.
\newblock \href {https://doi.org/10.1109/FOCS57990.2023.00057}
  {\path{doi:10.1109/FOCS57990.2023.00057}}.

\bibitem{FeldmanSS20}
Dan Feldman, Melanie Schmidt, and Christian Sohler.
\newblock Turning big data into tiny data: Constant-size coresets for k-means,
  pca, and projective clustering.
\newblock {\em {SIAM} J. Comput.}, 49(3):601--657, 2020.
\newblock \href {https://doi.org/10.1137/18M1209854}
  {\path{doi:10.1137/18M1209854}}.

\bibitem{FichtenbergerLN21}
Hendrik Fichtenberger, Silvio Lattanzi, Ashkan Norouzi{-}Fard, and Ola
  Svensson.
\newblock Consistent k-clustering for general metrics.
\newblock In D{\'{a}}niel Marx, editor, {\em Proceedings of the 2021 {ACM-SIAM}
  Symposium on Discrete Algorithms, {SODA} 2021, Virtual Conference, January 10
  - 13, 2021}, pages 2660--2678. {SIAM}, 2021.
\newblock \href {https://doi.org/10.1137/1.9781611976465.158}
  {\path{doi:10.1137/1.9781611976465.158}}.

\bibitem{GuK99}
Sudipto Guha and Samir Khuller.
\newblock Greedy strikes back: Improved facility location algorithms.
\newblock {\em J. Algorithms}, 31(1):228--248, 1999.
\newblock URL: \url{http://dx.doi.org/10.1006/jagm.1998.0993}, \href
  {https://doi.org/10.1006/jagm.1998.0993} {\path{doi:10.1006/jagm.1998.0993}}.

\bibitem{Har-PeledM04}
Sariel Har{-}Peled and Soham Mazumdar.
\newblock On coresets for k-means and k-median clustering.
\newblock In L{\'{a}}szl{\'{o}} Babai, editor, {\em Proceedings of the 36th
  Annual {ACM} Symposium on Theory of Computing, Chicago, IL, USA, June 13-16,
  2004}, pages 291--300. {ACM}, 2004.
\newblock \href {https://doi.org/10.1145/1007352.1007400}
  {\path{doi:10.1145/1007352.1007400}}.

\bibitem{HenzingerK20}
Monika Henzinger and Sagar Kale.
\newblock Fully-dynamic coresets.
\newblock In Fabrizio Grandoni, Grzegorz Herman, and Peter Sanders, editors,
  {\em 28th Annual European Symposium on Algorithms, {ESA} 2020, September 7-9,
  2020, Pisa, Italy (Virtual Conference)}, volume 173 of {\em LIPIcs}, pages
  57:1--57:21. Schloss Dagstuhl - Leibniz-Zentrum f{\"{u}}r Informatik, 2020.
\newblock URL: \url{https://doi.org/10.4230/LIPIcs.ESA.2020.57}, \href
  {https://doi.org/10.4230/LIPICS.ESA.2020.57}
  {\path{doi:10.4230/LIPICS.ESA.2020.57}}.

\bibitem{hu2018nearly}
Wei Hu, Zhao Song, Lin~F Yang, and Peilin Zhong.
\newblock Nearly optimal dynamic $ k $-means clustering for high-dimensional
  data.
\newblock {\em arXiv preprint arXiv:1802.00459}, 2018.

\bibitem{huang23optimal}
Lingxiao Huang, Jian Li, and Xuan Wu.
\newblock On optimal coreset construction for euclidean $(k,z)$-clustering.
\newblock {\em CoRR}, abs/2211.11923, 2022.
\newblock URL: \url{https://doi.org/10.48550/arXiv.2211.11923}, \href
  {http://arxiv.org/abs/2211.11923} {\path{arXiv:2211.11923}}, \href
  {https://doi.org/10.48550/ARXIV.2211.11923}
  {\path{doi:10.48550/ARXIV.2211.11923}}.

\bibitem{HuangV20}
Lingxiao Huang and Nisheeth~K. Vishnoi.
\newblock Coresets for clustering in euclidean spaces: importance sampling is
  nearly optimal.
\newblock In Konstantin Makarychev, Yury Makarychev, Madhur Tulsiani, Gautam
  Kamath, and Julia Chuzhoy, editors, {\em Proceedings of the 52nd Annual {ACM}
  {SIGACT} Symposium on Theory of Computing, {STOC} 2020, Chicago, IL, USA,
  June 22-26, 2020}, pages 1416--1429. {ACM}, 2020.
\newblock \href {https://doi.org/10.1145/3357713.3384296}
  {\path{doi:10.1145/3357713.3384296}}.

\bibitem{LattanziS19}
Silvio Lattanzi and Christian Sohler.
\newblock A better k-means++ algorithm via local search.
\newblock In Kamalika Chaudhuri and Ruslan Salakhutdinov, editors, {\em
  Proceedings of the 36th International Conference on Machine Learning, {ICML}
  2019, 9-15 June 2019, Long Beach, California, {USA}}, volume~97 of {\em
  Proceedings of Machine Learning Research}, pages 3662--3671. {PMLR}, 2019.
\newblock URL: \url{http://proceedings.mlr.press/v97/lattanzi19a.html}.

\bibitem{MakarychevMR19}
Konstantin Makarychev, Yury Makarychev, and Ilya~P. Razenshteyn.
\newblock Performance of johnson-lindenstrauss transform for \emph{k}-means and
  \emph{k}-medians clustering.
\newblock In Moses Charikar and Edith Cohen, editors, {\em Proceedings of the
  51st Annual {ACM} {SIGACT} Symposium on Theory of Computing, {STOC} 2019,
  Phoenix, AZ, USA, June 23-26, 2019}, pages 1027--1038. {ACM}, 2019.
\newblock \href {https://doi.org/10.1145/3313276.3316350}
  {\path{doi:10.1145/3313276.3316350}}.

\bibitem{MettuP04}
Ramgopal~R. Mettu and C.~Greg Plaxton.
\newblock Optimal time bounds for approximate clustering.
\newblock {\em Mach. Learn.}, 56(1-3):35--60, 2004.
\newblock URL: \url{https://doi.org/10.1023/B:MACH.0000033114.18632.e0}, \href
  {https://doi.org/10.1023/B:MACH.0000033114.18632.E0}
  {\path{doi:10.1023/B:MACH.0000033114.18632.E0}}.

\bibitem{SchwiegelshohnS22}
Chris Schwiegelshohn and Omar~Ali Sheikh{-}Omar.
\newblock An empirical evaluation of k-means coresets.
\newblock In Shiri Chechik, Gonzalo Navarro, Eva Rotenberg, and Grzegorz
  Herman, editors, {\em 30th Annual European Symposium on Algorithms, {ESA}
  2022, September 5-9, 2022, Berlin/Potsdam, Germany}, volume 244 of {\em
  LIPIcs}, pages 84:1--84:17. Schloss Dagstuhl - Leibniz-Zentrum f{\"{u}}r
  Informatik, 2022.
\newblock URL: \url{https://doi.org/10.4230/LIPIcs.ESA.2022.84}, \href
  {https://doi.org/10.4230/LIPICS.ESA.2022.84}
  {\path{doi:10.4230/LIPICS.ESA.2022.84}}.

\end{thebibliography}

\appendix
\section{Coreset via Uniform Sampling}
\label{sec:uniformCoreset}
We sketch in this section the proof of \Cref{lem:coresetStatic}. 
The proof is based on Bernstein centration inequality, and follows the line of the proof in \cite{stoc21}. We mention here the differences. Fix a set $S$ with $k$ centers: the goal is to show that the cost of $S$ is preserved with some very large probability, and then union-bound over all $S$s.

First, the points of $G$ are divided into \emph{types}: the \emph{tiny} types are all points with $\cost(p, S) \leq \eps/2 \cost(p, A)$; the \emph{interesting} type $\ell$ is all points with $2^\ell \cost(p, A) \leq \cost(p, S) \leq 2^{\ell+1} \cost(p, A)$, for $\log \eps/2 \leq \ell \leq 2\log (8/\eps) $. Finally, the remaining points with $\cost(p, S) \geq (8/\eps)^2 \cost(p, A)$ are the \emph{huge} points.

The tiny points are so cheap that they contribute at most $\eps / 2 \cost(G, A)$ both in $G$ and in $\coreset$: the proof of Lemma 5 in \cite{stoc21} applies directly to our sampling distribution (we refer to the numbering of the full version on arxiv).

For the huge points, \cite{stoc21} shows that it is enough for the coreset to preserve approximately the mass of each cluster of $A$ (which they call event $\calE$). 
Call $C_1, ..., C_k$ the clusters of $A$. In our setting, because each  of the $\delta$ coreset point $p$ is weighted $w(p) |G|/\delta$, this event corresponds to: 
\begin{equation}
    \label{eq:eventE}
    \frac{|G|}{\delta} \cdot \sum_{p \in C_i \cap \coreset} w(p) = (1\pm \eps) \sum_{p \in C_i} w(p). 
\end{equation}

To prove this equation, note that each sampled point is part of $C_i$ with probability $|C_i|/|G| \geq 1/(8k)$ -- from \Cref{prop:groups}. Therefore, the proof of Lemma 6 of \cite{stoc21} can be directly adapted.

Note that, in particular, Equation~\ref{eq:eventE} shows that the total weight of the coreset is almost equal to the weight of $G$, as claimed in \Cref{lem:coresetStatic}.

Second, Lemma 7 of \cite{stoc21} shows that, given Equation~\ref{eq:eventE} holds, the points in clusters that are intersecting with huge type have cost preserved by the coreset. 
Therefore, it only remains to deal with the interesting points, in clusters that contain no points from the huge type (called $L_S$ in \cite{stoc21}).

Note that due to the choice of weights, $\cost(L_S \cap \coreset, S)$ is an unbiased estimator of $\cost(L_S \cap P, S)$. Further, $\cost(L_S \cap \coreset, S)$ can be expressed as a sum of $|\coreset|$ independent random variables: let $X_i$ be the variable equal to $w(p) \cdot \frac{|G|}{\delta} \cost(p, S)$ if $p$ is the $i$-th sampled point, and $p \in L_S$. If $p \notin L_S$, $X_i$ is equal to $0$.

The key argument in the proof of \cite{stoc21} is to bound the variance of $X_i$. In our case, we have:

\begin{align*}
    \E[X_i^2] &= \sum_{p \in L_S} \lpar w(p) \cdot \frac{|G|}{\delta} \cost(p, S) \rpar^2 \Pr[p\text{ is the $i$-th sampled point}]\\
    &= \sum_{p \in L_S} \lpar w(p) \cdot \frac{|G|}{\delta} \cost(p, S) \rpar^2 \cdot \frac{1}{|G|}\\
    &\leq \frac{4|G|}{\delta^2} \sum_{p \in L_S} \cost(p, S) ^2,
\end{align*}
where we used that for any point $w(p) \in [1,2]$. Now, for $p \in L_S$, we have $\cost(p, S) \leq (8/\eps)^2 \cost(p, A)$. Further, since all points in the group $G$ have same cost, up to a factor 2 (from \Cref{prop:groups}), $\cost(p, A) \leq 2\cost(G, A)/ |G|$. 
Therefore, we get:
\begin{align*}
    \E[X_i^2] &\leq \frac{4|G|}{\delta^2} \sum_{p \in L_S} \cost(p, S) \cdot \frac{64}{\eps^2} \cdot \frac{2\cost(G,A)}{|G|}\\
    &\leq \frac{512}{\delta^2 \eps^2} \cost(G, A) \cost(G, S).
\end{align*}

Using this bound in Lemma 8 of \cite{stoc21} concludes the proof for interesting types. 
Putting all the results together and performing a union-bound does not depend on the sampling distribution, and therefore those results are enough to finish the proof of \Cref{lem:coresetStatic}.

This however shows the slightly suboptimal result that $\coresetSize = O(k \log(n) \eps^{-4})$.
To reduce the dependency in $\eps$ to $\eps^{-2}$, \cite{stoc21} use a technique to reduce the variance of the estimator (see their section 5.6): this works the same way as what we described above, and conclude \Cref{lem:coresetStatic}.

\end{document}